\documentclass[utf8
               ]{jetp} 

\usepackage{cuted}

\usepackage{capt-of}

\usepackage{multicol}

\usepackage{graphicx}
\usepackage{tikz}
\usepackage{pgfplots}
\usepackage{systeme}
\usepackage{cite}
\usepackage{amsmath,amsfonts,amssymb,amsthm,mathtools} 
\usepackage{verbatim} 
\usepackage{graphbox}

\begin{document}
\english
\title{Interference of a chain of Bose condensates in the Pitaevskii-Gross approximation
}
\setaffiliation1{Department of Physics,  Lomonosov Moscow State University, 119991, Moscow, Russia}
\setaffiliation2{International Center for Quantum Optics and Quantum Technologies LLC,  121205, Moscow, Russia}
\setaffiliation3{A.V. Gaponov-Grekhov Institute of Applied Physics of the Russian Academy of Sciences,  603950, Nizhny Novgorod,  Russia}
\setaffiliation4{Moscow Institute of Physics and Technology, 141701, Dolgoprudny, Moscow region, Russia}

\setauthor{I.~N.}{Mosaki}{12}
\setauthor{A.~V.}{Turlapov}{234}\email{a\_turlapov@mail.ru}

\rtitle{Interference of a chain of Bose condensates in the Pitaevskii-Gross approximation}
\rauthor{I.~N.  Mosaki,  A.~V. Turlapov}

\abstract{
A long chain of Bose condensates freely expands and interferes after being released from an optical lattice. The interference fringes are well resolved both in the case of equal phases of the condensates and in the case of fluctuating phases. In the second case the positions of the fringes also fluctuate. The spectrum of the spatial density distribution, however, is reproducible despite the fluctuations. Moreover two types of peaks are distinguishable in the spectrum. The first type arises due to the phase fluctuations, the second type is associated with the coherence between the condensates. In the framework of the Pitaevskii-Gross equation we calculate the interference of the condensates and compare the calculation with experiment [Phys. Rev. Lett. 122, 090403 (2019)]. The calculation reproduces the positions of the spectrum peaks, including the dependence on the interparticle interaction.  The calculated heights of the peaks, however, in some cases differ with the experimental ones.
}

\maketitle



\begin{multicols}{2}

\section{Introduction}

In optics,  when light passes through a material structure,  diffraction and interference occur.
In atomic physics,   matter and light switch roles, and interference of de Broglie waves  emerges  with an initial condition formed by light.
The basis for  this role reversal was laid  by Kapitza and Dirac,  who predicted  the diffraction of electrons on a standing light wave \cite{KapitzaDirac1933}.
For atoms,  a short exposure to a near-resonant standing light wave  
has been predicted to produce a
 periodic spatial modulation of the wave function \cite{KazantsevBeamSplit1980rus}. 
The diffraction of  atomic de Broglie waves resulting from such modulation has been observed \cite{PritchardThinSWDiffraction1986},   is frequently utilized, and is traditionally referred to as the Kapitza-Dirac diffraction.

The Talbot effect~\cite{Talbot}  is a typical example of the resemblance between the interference of  light and de Broglie waves.
In optics,   the Talbot effect  manifests as the self-imaging of a long chain of coherent sources at a certain,  not very large,  distance.
Analogues of the optical Talbot effect  were found in various systems, including   acoustic waves~\cite{AcousticTalbot1985,USoundTalbot2017rus},  plasmons~\cite{PlasmonTalbot2009}, spin waves~\cite{SpinwaveTalbot2012}, polaritons  \cite{Gao2016}, electron de Broglie waves~\cite{DenisovTalbot1996,DenisovTalbot1999}, and atoms~\cite{PritchardTalbot1995}. 
In quantum systems,  a necessary condition for  the self-imaging is a spatial periodicity of the wave function,  $\psi(z+d)=\psi(z)$,  or its analogue.
For  atoms and  molecules,   the Talbot effect can   appear  as a particular case of the Kapitza-Dirac diffraction \cite{PhillipsTalbot1999,FluctPhaseTalbot2017,ChineeseTalbot2024}.
To apply the  initial modulation,  atoms do not have to pass through a standing wave; it is sufficient to  turn on the light for a short time, leaving the atoms motionless.
The self-imaging of the wave function $\psi(z)$ will occur at the same location after the Talbot time $T_d=md^2/(\pi\hbar)$, where $m$ is the particle mass.


The role reversal between  light and  matter in interference has enhanced the understanding of the Talbot effect.
As an initial condition,  consider  a chain of Bose-Einstein condensates prepared in an optical lattice, 
where the condensates are trapped in  the antinodes of the standing light wave,  as  depicted in Fig. \ref{fig:InitialCondition}a. 
\begingroup
\hspace{-0.6cm}
\begin{minipage}{1.0\linewidth}
	\includegraphics[width=1.0\textwidth]{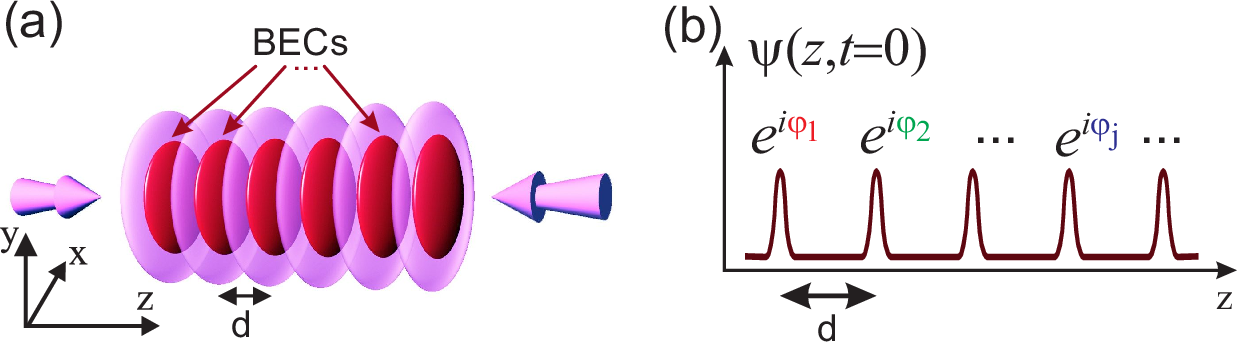}
	\captionof{figure}{
		a)  Bose-Einstein condensates 
		(BECs)
		 in the optical lattice  prior to the release and interference.
		The condensates are shown in dark red, the standing-wave intensity is shown  in light purple.
		b) The initial wave function of the condensates~--- its module is periodic,  and the phase differences between neighbors are in the general case arbitrary.
		\label{fig:InitialCondition}
	}
\end{minipage}
\endgroup
At  time $t=0$,  the light is 
turned off,  
causing the condensates to expand and interfere in the free space.
If the condensates are coherent, the initial chain of the condensates is restored at $t=T_d$, 
demonstrating the Talbot effect \cite{NaegerlTalbot2011,RandomPhaseInteference2019}.
The coherence between the condensates  can be reduced by increasing the lattice depth or the temperature \cite{DalibardBECInterference2004,RandomPhaseInteference2019}, while  
the coherence within each individual condensate is maintained.
The interference fringes are well resolved even if the phases of the adjacent condensates are completely random.
 In the latter case,  
 the positions and intensities of the fringes,  as well as the distance between them,  
  fluctuate from one repetition of the experiment to another.
  However,  the spectrum of the spatial density distribution  is reproducible despite the fluctuations.
The spectrum consists of  equidistant peaks,  indicating spatial order,  although with a period different from the initial one,  $d$ \cite{RandomPhaseInteference2019}.
In  classical optics,  such spatial order in the interference of a chain of elements with  fluctuating phases  
 has not been observed.
When  partial coherence is present among the condensates, 
the spectrum exhibits two distinct types of peaks: 
one type arises due to the phase fluctuations,  while the other type is associated with the coherence between the condensates.

In this paper,  we model the interference of a long chain of  Bose condensates released from an optical lattice.
We calculate the interference  for arbitrary coherence between the condensates.
 The model is based on the Pitaevskii-Gross equation \cite{PitaevskiiGPE1961Rus,GrossGPE1961}.
 We compare the calculations with the experimental results \cite{RandomPhaseInteference2019}.
 The calculation quantitatively reproduces   the positions of the spectrum peaks,  including  shifts   caused by the interatomic interactions.
 However, in some cases, the peak amplitudes differ from the experimental ones.

The model and the approximations are described in Section \ref{sec:Model}.
General properties of the interference fringes and the corresponding spatial spectrum are discussed in Section \ref{sec:Spectrum}.
The comparison between  the calculation and the experiment \cite{RandomPhaseInteference2019}  is presented in Section \ref{sec:Compare}. 
The conclusion is provided in Section  \ref{sec:Conclusion}.

\section{Model and initial conditions}\label{sec:Model}

We  choose a model and  initial conditions in accordance with the experiment  \cite{RandomPhaseInteference2019}.
In this experiment,   condensates were produced from a gas of bosonic molecules $^6$Li$_2$. 
Initially,  the condensates were trapped in an optical lattice (Fig. \ref{fig:InitialCondition}a) with the potential
\begin{equation}
	V_s(\mathbf{r}) = sE_{\text{rec}} \left(1 -  e^{-\frac{2mE_{\text{rec}} \rho^2 }{ ( \hbar \lambda )^2}} \cos^2 \left( \frac{\pi z}{d} \right)  \right),
	\label{latt_potential}
\end{equation}
where $s$ is the dimensionless lattice depth, 
$\lambda=27.4$ is the anisotropy parameter of the disk shaped traps,
$m$ is the boson mass, 
 $\rho\equiv\sqrt{x^2+y^2}$,  and $E_{\text{rec}}=\hbar^2\pi^2/(2md^2)$ is the lattice photon recoil energy. 
 The harmonic expansion of the potential \eqref{latt_potential} 
 near each minimum
 gives frequencies of the microtraps $\omega_z=2\sqrt sE_{\text{rec}}/\hbar$ and $\omega_\perp= \omega_z/\lambda$.

In the experiment,  
some of the peaks 
in the spectrum were shifted relative to the calculations for the non-interacting particles. 
Therefore,   the model should incorporate interparticle interactions.
 Considering only the s-wave interaction between the bosons,   the dynamics is described by the Pitaevskii-Gross equation  \cite{PitaevskiiGPE1961Rus,GrossGPE1961} for the wave function of the condensate $\Psi(\mathbf{r},t)$:
\begin{equation}
i\hbar \frac{\partial\Psi}{\partial t} = -\frac{\hbar^2}{2m}\nabla^2\Psi + g\left|\Psi\right|^2\Psi, 
	\label{GP}
\end{equation}
The equation is valid because the interactions are small~--- $n_{\max}^{1/3}a\lesssim0.1$,  where $n_{\max}$ is the density in the cloud center.

When
choosing the initial condition $\Psi(\mathbf{r},t=0)$,  we take into account that the lattice is  sufficiently
 deep,  and the condensates do not overlap.
Under the typical experimental conditions,  $\mu/(\hbar\omega_z)=0.3$,  where $\mu$ is the chemical potential of bosons measured from $\hbar\omega_z/2$. 
Hence,  the gas is nearly kinematically two-dimensional, 
with most of the bosons  in the lowest state of the harmonic oscillator along $z$.
Therefore,   $\Psi(\mathbf{r},0)$ can be represented as a product of the longitudinal and radial parts: $\Psi(\mathbf{r})=\psi(z,0)\chi(\rho,0)$.

We choose the longitudinal part $\psi(z,0)$ 
as a  sum of  gaussians with their own phases,  as shown in Fig.  \ref{fig:InitialCondition}b:
\begin{equation}
\psi(z,t=0)=\frac{N_{0}^{1/2}}{(2\pi)^{1/4}\sigma^{1/2}}\sum\limits_{j=1}^Ke^{-\frac{{(z-jd)}^2}{4\sigma^2}}e^{i\varphi_j},  
	\label{initial}
\end{equation}
where $N_{0}$ is the number of the bosons in each condensate,  $K$ is the number of the condensates in the chain. 
The gaussian shape follows from the closeness to the two-dimensionality, while the individual phases $\varphi_j$ are well-defined due to the localization.
We assume that
the phase differences between the neighbouring condensates $\varphi_j-\varphi_{j+1}$ 
are normally distributed random variables.
Then the phase correlation decays exponentionally,  similar to the thermal fluctuations \cite{BoseChainFluctPitaevskii2001}: 
$\langle\cos(\varphi_j-\varphi_l)\rangle=\alpha^{|j-l|}$,
  where 
  $\alpha=\langle\cos(\varphi_j-\varphi_{j+1})\rangle\simeq\overline{\cos(\varphi_j-\varphi_{j+1})}$
  is the coherence factor, 
$\langle...\rangle$ denotes averaging over experiment repetitions or calculations with random sets of phases $\{\varphi_j\}$ with  the identical   dispersion,   and the overline  indicates  averaging over the chain elements.

Since $\mu/(\hbar\omega_\perp)\simeq9\gg1$,  the radial part $\chi(\rho,0)$ can be estimated via the Thomas-Fermi approximation:
\begin{equation}
	\chi(\rho,  t =0)   = \frac1{R_{\text{TF}}}\sqrt{\frac2\pi\left(1-\frac{\rho^2}{R_{\text{TF}}^2}\right)}
	\label{radial_wave_func}
\end{equation}
if $\rho<R_{\text{TF}}$,  and 0 if $\rho>R_{\text{TF}}=2l_{\perp}(2N_0a/(\sqrt\pi\sigma))^{1/4}$, 
where $R$ is the Thomas-Fermi radius,  and $l_{\perp}=\sqrt{\hbar/(2m\omega_\perp)}$ is the radial oscillator length.

The radial and axial sizes of the condensates,   $R_{\text{TF}}$ and $\sigma$,  are 
interrelated.
The axial size,  $\sigma$,   can exceed
the RMS oscillator lenght $l_z=\sqrt{\hbar/(2m\omega_z)}$ due to the interactions.
Therefore,  $R_{\text{TF}}$ and $\sigma$  must be determined
self-consistently. To achieve  this,
we substitute $\Psi(\mathbf{r},0) = \psi(z,0)\chi(\rho,0)$ into the Pitaevskii--Gross energy functional
\begin{equation}
\int\left( 
\frac{\hbar^2}{2m}|\nabla\Psi|^2 + \frac{m(\omega_z^2z^2+\omega_\perp^2\rho^2)}2|\Psi|^2 + 
\frac g2|\Psi|^4  
\right) d^3
\textbf r
\end{equation}
and minimize it with respect to $\sigma$. 
The condition of the minimum
\begin{equation}
\left(\frac\sigma{l_z}\right)^4 - \frac{(8al_z^3N_0)^{1/2}}{3\pi^{1/4}l_\perp^2} \left(\frac\sigma{l_z}\right)^{3/2} - 1=0 
	\label{broadening}
\end{equation}
 leads to the value $\sigma/l_z=1.3$--$1.5$ under the experimental conditions \cite{RandomPhaseInteference2019}.

Due to the strong anisotropy  of the condensates the expansion in the radial direction is much slower than in the $z$-direction. Therefore,  we can neglect the radial expansion.
Substituting
$ \Psi(\mathbf{r},t)=\psi(z,t)\chi(\rho,t=0)$
into the  equation  \eqref{GP}
and   averaging over the radial coordinate yields the equation for $\psi(z,t)$:
\begin{equation}  
i\hbar\frac{\partial\psi}{\partial t} = -\frac{\hbar^2}{2m}\frac{\partial^2\psi}{\partial z^2} +  \frac{4\pi^{1/4}}3\left(\frac{2a\sigma}{N_0}\right)^{1/2}\hbar\omega_{\perp}|\psi|^2\psi.
     \label{GP1D}
\end{equation}  
The equation (\ref{GP1D}),  the initial condition (\ref{initial}),  and the 
 equation for the axial condensate size (\ref{broadening})
  constitute the model  used in the subsequent calculations.
The purpose of the modelling is to compute the column density $n_2(x,z)$ at time  $t$,  as this value is measurable in experiments and directly indicates the result of interference.
The column density  is  derived from the local density $n(\textbf r)=|\Psi(\textbf r)|^2$  by integration:  $n_2(x,z)=\int n(\textbf r)dy$.

\section{General properties of interference fringes and their spatial spectrum}
\label{sec:Spectrum}

We now demonstrate the general properties of the interference fringes at time $t$ after the abrupt turning off of the optical lattice at $t=0$ and the subsequent evolution of the condensates in the free space.

If there are no interactions
and the phases of the condensates are equal,   the initial density distribution is exactly restored after the Talbot time $T_d$:  $\psi(z,0)=\psi(z,T_d)$. 
The presence of the  interparticle interactions disrupts this exact time periodicity;  however,  at $t=T_d$,  the density profile is nonetheless close to the initial one. This is illustrated in Figs. 2a,b, which show the calculated column density at $t=0$ and $T_d$. 
    \begingroup
\hspace{-0.6cm}
\begin{minipage}{1.0\linewidth}
	\hspace{0.3cm}
	\begin{minipage}{0.8\linewidth}
		\includegraphics[scale=0.25]{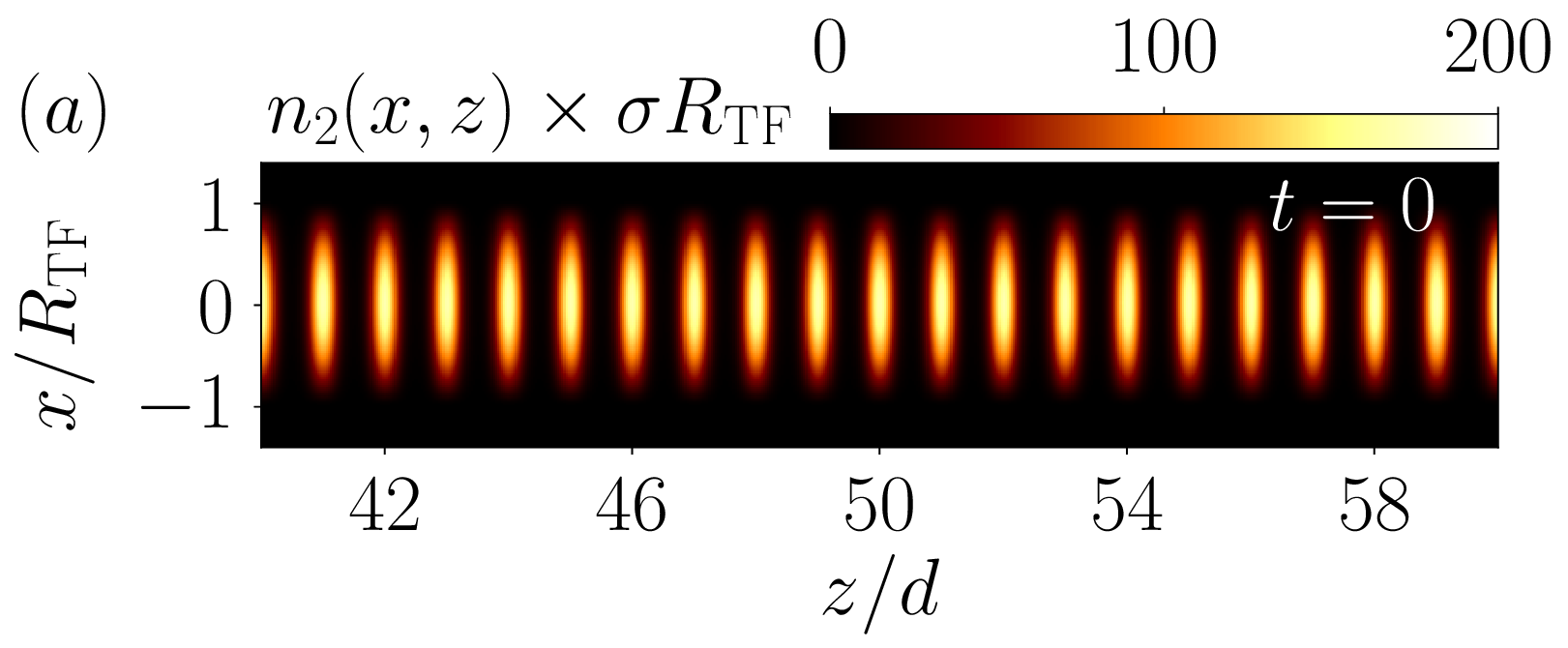}
		
		\vspace{-0.3cm}
		\includegraphics[scale=0.25]{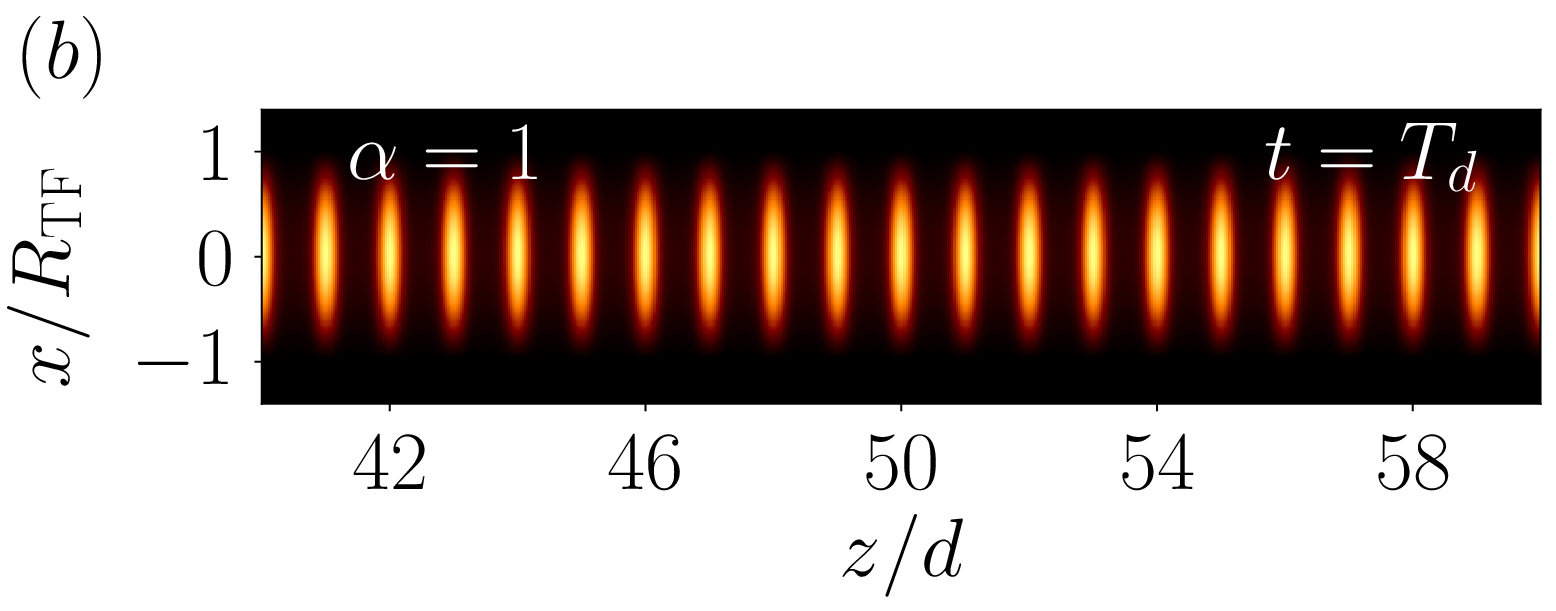}
		
		\vspace{-0.3cm}
		\includegraphics[scale=0.25]{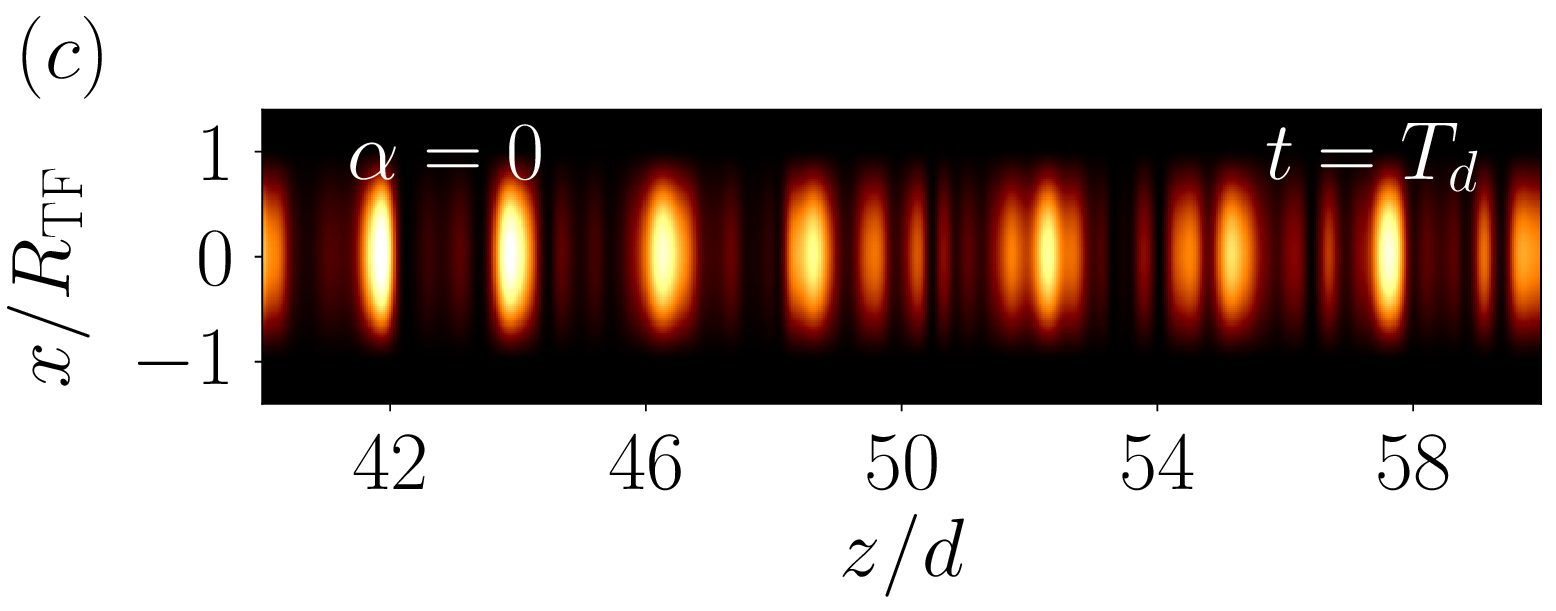}
		
	\end{minipage}
	\vspace{-0.2cm}
	\captionof{figure}{
	Numerically calculated column density of a chain of  condensates $n_2(x,z)$: 
(a) in the optical lattice at $t=0$, 
(b) at $t=T_d$ for the identical initial phases of the condensates ($\alpha=1$), 
(c) at $t=T_d$ for the completely disordered initial phases of the condensates ($\alpha=0$).
		\label{density2d}
	}
\end{minipage}
\endgroup

\;

The disorder of the  initial phases of the condensates  significantly alters
the positions of the interference fringes.
Fig. \ref{density2d}c shows an example of the numerically calculated fringe pattern for completely disordered initial phases.
Although the sharp fringes  are preserved,
 their positions now depend on the specific values of the initial phases.
As a result, the positions of the fringes change with each repetition of the experiment.

\begin{figure*}
	\begin{minipage}{1.0\linewidth}
		\begin{center}
			\includegraphics[scale=0.8]{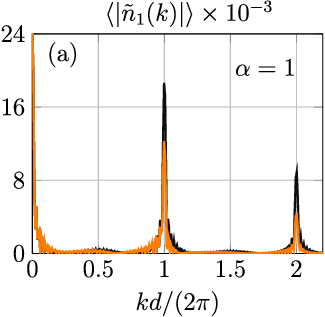}
			\hspace{0.7cm}
			\includegraphics[scale=0.8]{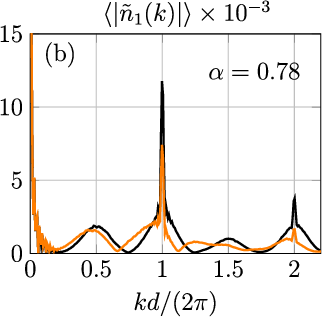}
			\hspace{0.7cm}
			\includegraphics[scale=0.8]{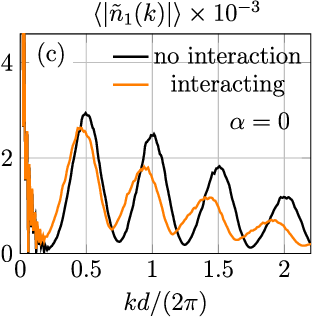}
		\end{center}
		\vspace{-0.7cm}
		\captionof{figure}{ 
			\label{spec2}
	Average absolute value 
	of interference fringe spectrum $\langle|\tilde{n}_1(k,T_d)|\rangle$ at  $t=T_d$
	for different degrees of initial phase disorder:
(a) $\alpha=1$~--- the phases are identical; 
(b) $\alpha=0.78$~--- the phases are partially disordered;
 (c)  $\alpha=0$~--- the phases are completely disordered.
The calculations for the  interacting and non-interacting gas are shown.
		}
	\end{minipage}
\end{figure*}

The regularity in the interference fringes with disordered phases can be revealed through the analysis of their spectrum
\begin{equation}
\tilde n_1(k,t)=\int n_1(z,t)e^{-ikz}dz \label{eq:ModelSpectrum}
\end{equation}
with the linear density 
$n_1(z,t)=\int n_2(x,z,t)dx$.
Within the model  
$\tilde n_1(k,t)=\int|\psi(z,t)|^2e^{-ikz}dz$.
Figure \ref{spec2} shows the numerically calculated spectra $\langle|\tilde n_1(k,T_d)|\rangle$ for different degrees of condensate phase correlation.

The calculations were performed using parameters  typical for the experiment \cite{RandomPhaseInteference2019}: $N_0=500$, $a = 1520$ bohr, $s=20$, $K=50$, $d=5.3$~$\mu$m,  which give $\sigma = 1.36 l_{z}$.
Additionally,  the calculation for  $a=0$ was provided.
To reduce the small-scale noise,  the spectra were averaged over 800 repetitions.
In all cases,  the spectra exhibit peaks, indicating the spatial order in the interference fringes.

When the phases are equal,  the spatial spectrum consist of the narrow peaks at $k=2\pi l/d$, $l\in\mathbb{Z}$ as shown in Fig. \ref{spec2}a. 
For the completely disordered condensates $(\alpha=0)$ and $a=0$,  it was shown earlier analytically \cite{RandomPhaseInteference2019} that the spectrum again consists of the equidistant peaks, but these peaks are wider and appear at different momenta $k=\pi lT_d/(td)$. 
The numerically calculated result, shown in Fig. 3c by the black curve,  closely aligns with this analytical prediction. For the partially disordered phases the spectrum is depicted in Fig. \ref{spec2}b.  This spectrum features two types of peaks:
the narrow peaks,  which  correspond  to the Talbot effect,  and the wide peaks,     which arise due to the phase fluctuations.
The relative contributions of these two types of  peaks can be used for measuring the coherence factor $\alpha$.


The influence of the interparticle interactions on the two  types of peaks  is different,   as seen by comparing the black and orange curves in Fig.  \ref{spec2}. 
The positions of the narrow peaks are unaffected by the mean field.
In contrast, the wide peaks are shifted to lower momenta due to the mean field.

\section{Comparison with experiment}
\label{sec:Compare}

\begin{figure*}
	\begin{minipage}{1.0\linewidth}
		\hspace{-0.8cm}
		\begin{minipage}{0.5\linewidth}
			
			\vspace{-3.0cm}

			\large \hspace{2.0cm} \textbf{ Nearly equal phases}  \normalsize 
			
			\vspace{0.3cm}
			
			\hspace{0.05cm}
			\includegraphics[height=0.96in]{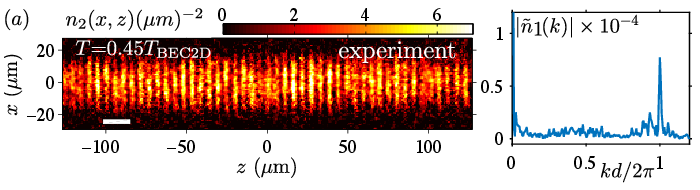}

			\vspace{0.0cm}
			\hspace{0.2cm}
			\includegraphics[height=1.07in]{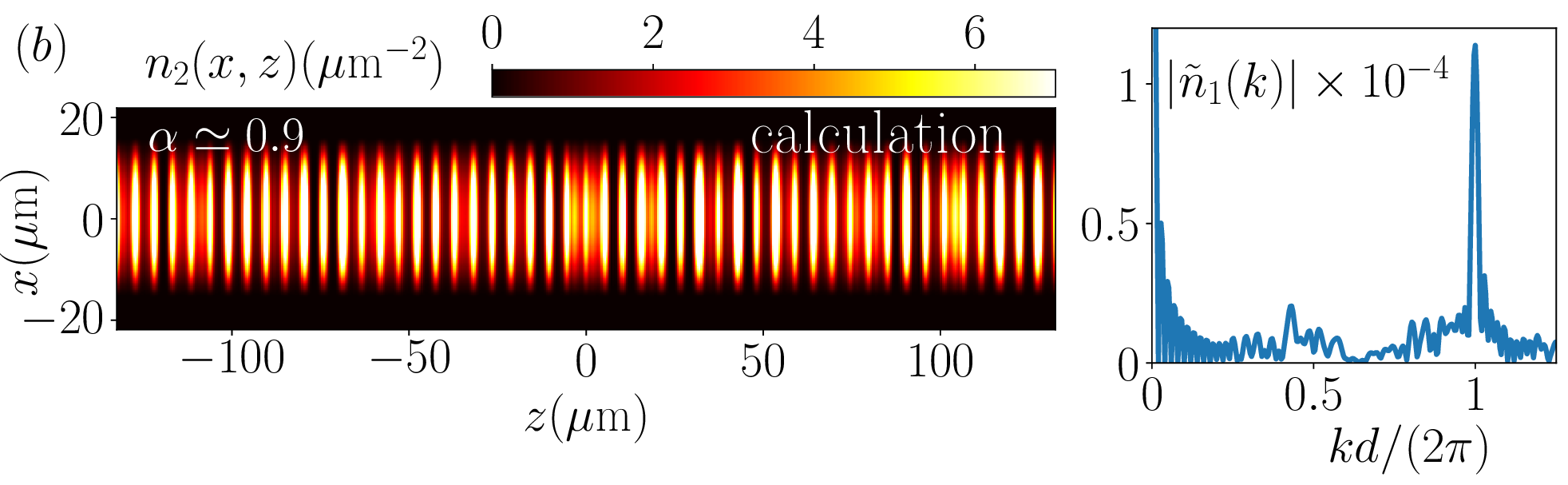}
		\end{minipage}
		\hspace{0.5cm}
		\begin{minipage}{0.5\linewidth}

			\vspace{-0.1cm}
			\large  \hspace{2cm} \textbf{Randomized phases} \normalsize 
			
			\vspace{0.3cm}
			\includegraphics[height=0.964in]{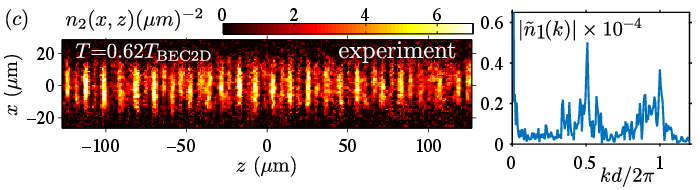}
			
			\vspace{0.05cm}
			\hspace{-0.05cm}
			\includegraphics[height=1.04in]{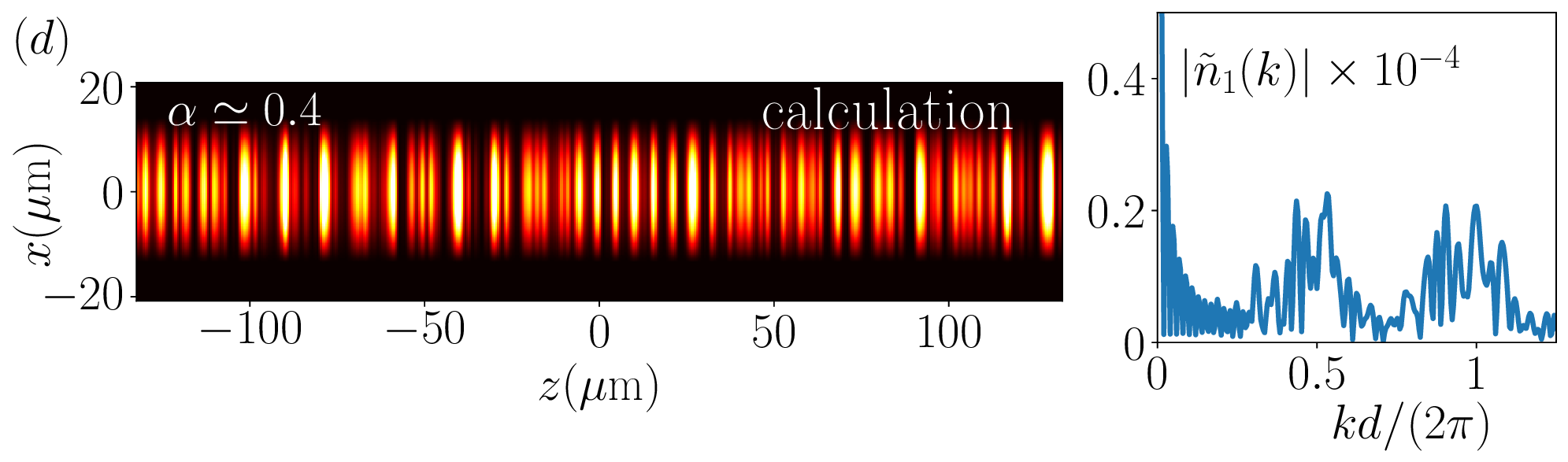}
			
			\vspace{0.2cm}
			\hspace{-0.04cm}
			\includegraphics[height=1.04in]{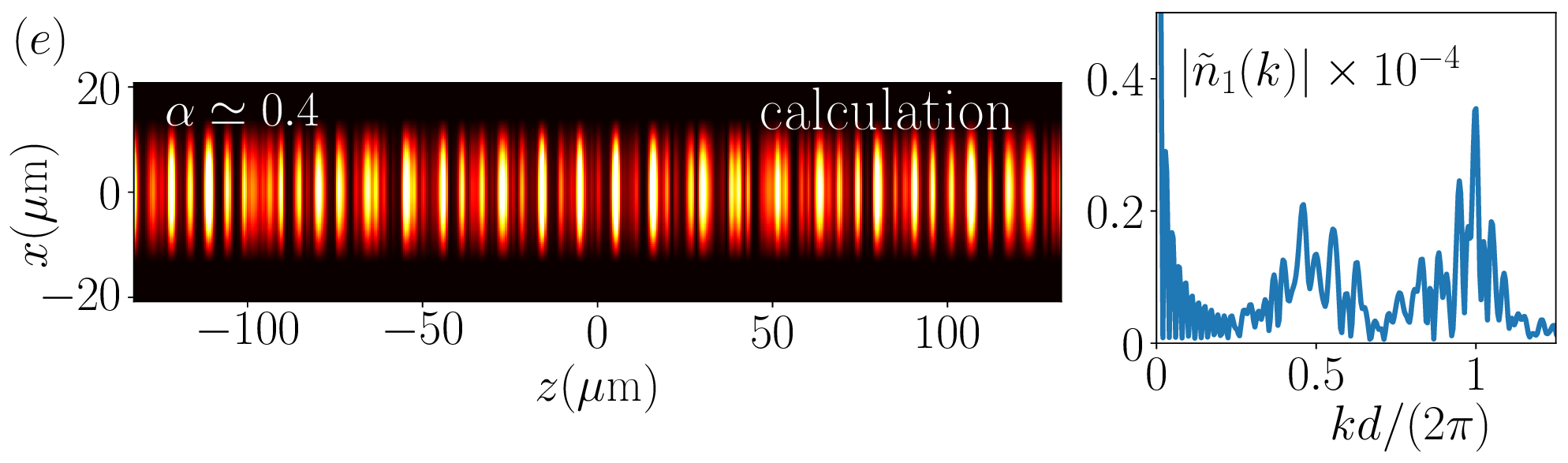}
		\end{minipage}
	\end{minipage}       
	\captionof{figure}{
		\label{experiment_photos}
(a), (c) Interference images
from the experiment \cite{RandomPhaseInteference2019} at $t=T_d$ and their spectra for a chain of  Bose condensates at temperatures $T=0.45T_{\text{BEC2D}}$ and $T=0.62T_{\text{BEC2D}}$,  respectively. 
(b) Simulation of the data from  fig. (a).
(d), (e) Simulation of the data from fig. (c),  using two different sets of phases $\{\varphi_j\}$ with the same coherence factor $\alpha$.
}
\end{figure*}

In the experiment \cite{RandomPhaseInteference2019},  the column density of the gas  was observed via the absorptive imaging.
Examples are shown in Figs.  \ref{experiment_photos}a,c,  which replicate Figs.  2b and 3b from \cite{RandomPhaseInteference2019}.

An image can be taken at any time $t$,  resulting in the destruction of the quantum state.
To obtain a new image, the experiment must be repeated.
In images  \ref{experiment_photos}a,c,  which were taken at  $t=T_d$,   the interference fringes are parallel,  indicating  the absence of  phase fluctuations within the condensates and justifying the introduction of a single phase $\varphi_j$ for the  $j$-th microcondensate.
From the experimental column densities the spectra were calculated.
These spectra are also shown in Figs. \ref{experiment_photos}a,c.    
The decrease in the coherence between the adjacent condensates and the crossover from the Talbot effect to the pronounced manifestations of the incoherence in the spectrum were achieved  by increasing the temperature $T$.
In Figs. \ref{experiment_photos}a,c,  the temperature  is marked in  units of the critical temperature of the two-dimensional Bose gas $T_{\text{BEC2D}}=\hbar\omega_{\perp}\sqrt{6N}/\pi$. 
The temperature was determined by fitting the radial density distribution at $t=0$ with a bimodal distribution.
The fitting also provided   $N_0/N$,   where $N$ is the total number of particles in one disk-shaped trap,  including  noncondensed particles.


The simulation of the data from Fig.  \ref{experiment_photos}a is shown in Fig.  \ref{experiment_photos}b,  the simulation of the data from \ref{experiment_photos}c is shown in Figs. \ref{experiment_photos}d,e.  
In the simulations,  we  used
$K=50$, $s=23.3$, $a=1520$ bohr, $d = 5.3$ $\mu$m,  $N_0=463$ for Fig.  \ref{experiment_photos}c,  and $N_0=271$ for Figs. \ref{experiment_photos}d,e.
The only adjustable parameter of the model was the coherence factor $\alpha$.
The phases in each set $\{\varphi_j\}$ were chosen randomly with the condition $\overline{\cos(\varphi_j-\varphi_{j+1})}=\alpha$.
By selecting an appropriate  $\alpha$,  we achieve the resemblance between the experimental and calculated spectra.
Exact reproduction of the interference fringes is not necessary for achieving the resemblance between the spectra.

\begin{figure*}
	\begin{minipage}{1.0\linewidth}
		\center
		\includegraphics[scale=0.8]{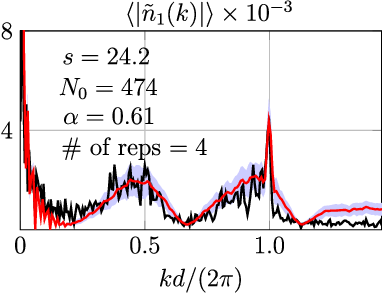}
		\hspace{0.5cm}
		\includegraphics[scale=0.8]{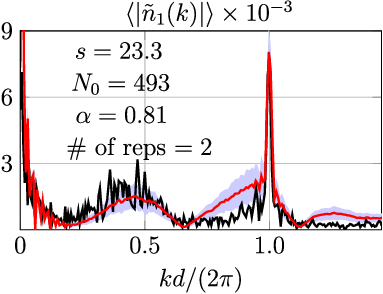}
		\hspace{0.5cm}
		\includegraphics[scale=0.8]{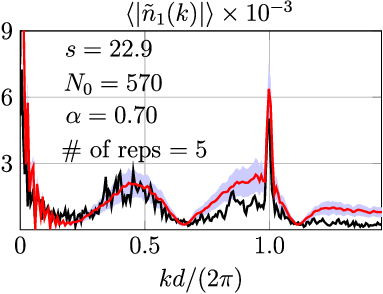}
		
		\vspace{0.5cm}
		\includegraphics[scale=0.8]{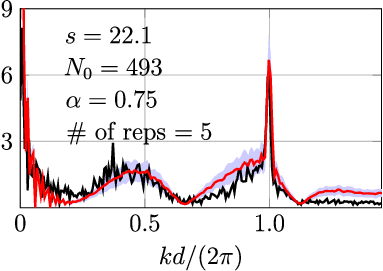}
		\hspace{0.3cm}
		\includegraphics[scale=0.8]{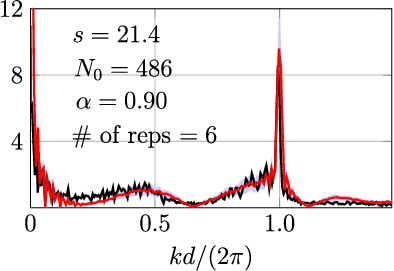}
		\hspace{0.3cm}
		\includegraphics[scale=0.8]{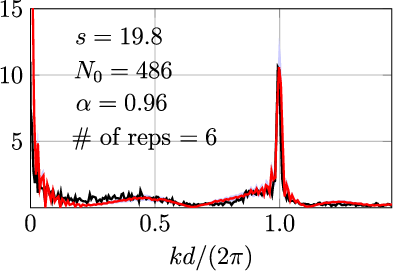}
		
		\vspace{0.5cm}
		\includegraphics[scale=0.8]{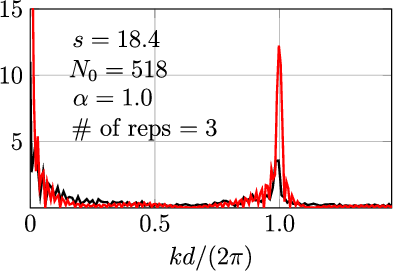}
		\hspace{0.3cm}
		\includegraphics[scale=0.8]{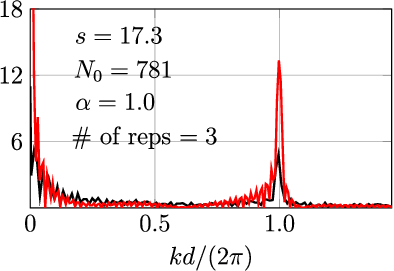}
		\hspace{0.3cm}
		\includegraphics[scale=0.8]{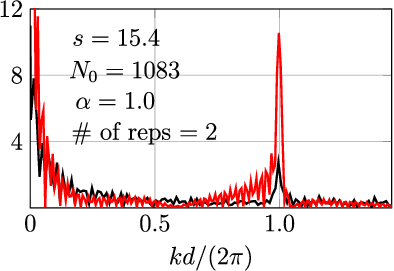}
		
		\vspace{0.5cm}
		\hspace{3.8cm}
		\includegraphics[scale=0.8]{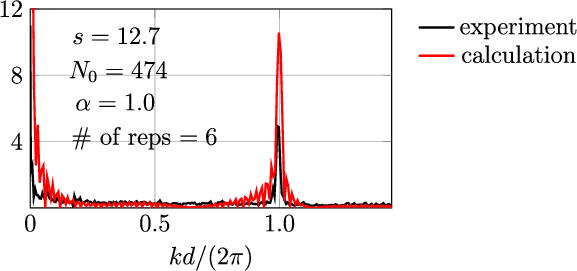}
		\vspace{-0.3cm}
		\captionof{figure}{\label{differences1}
The spectra of linear density distributions obtained after the free evolution of condensates during the Talbot period  $T_d$. 
The results of the experiment \cite{RandomPhaseInteference2019} and the calculations are shown.
The variable parameters are indicated in the figures,  <<\# of reps>> denotes the number of experimental repetitions.
The calculations are averaged over 200 sets $\{\varphi_j\}$,  the borders of the purple shading correspond to $\pm$ one standard deviation.
		}
	\end{minipage}
\end{figure*}

 Both the experimental and calculated spectra in Fig. 4 exhibit small-scale noise. 
In the case of the disordered phases,   the noise makes it difficult to determine whether the sharp peak at $kd/(2\pi)=1$ is associated with the Talbot effect or is a noise outlier.
This complication is clearly evident when comparing the spectra in Figs. \ref{experiment_photos}d and \ref{experiment_photos}e,  which are calculated with the same $\alpha$ but the different  sets of phases $\{\varphi_j\}$.
The small-scale noise can be suppressed by averaging over repetitions of the experiment or the calculation,  as shown in Figs.   \ref{spec2} and \ref{differences1}.

 \newpage

 The  crossover from the interference with the disordered phases to the Talbot effect can be observed in the data and calculations presented in Fig.  \ref{differences1}.
The experimental data are taken from Fig. 6 of the Supplemental Material of the paper   \cite{RandomPhaseInteference2019}. 
The result of the interference at $t=T_d$ is shown.
Both the experimental data and the simulations are averaged over  multiple repetitions.
The purple shading indicates the  magnitude of the  small-scale noise obtained in the calculations,   with the shading's borders representing   $\pm$ one standard deviation.
Due to the averaging,  the peak associated with the Talbot interference can be correctly identified and used to determine $\alpha$.
The coherence factor $\alpha$ depends on temperature \cite{BoseChainFluctPitaevskii2001}, providing a means for thermometry,  including the temperatures significantly below the critical temperature, which cannot be determined by the conventional bimodal fitting \cite{OberthalerNoiseThermometry2006,TalbotReview2019rus}.


 The crossover between the two regimes of the interference shown in Fig.  \ref{differences1} was achieved by gradually varying  the lattice depth $s$.
For  depths $s\leq18.4$,   the height of the calculated spectrum  clearly exceeds  the  experimental one.
The cause of this  discrepancy is not clear.
It is worth noting that  the Pitaevskii-Gross approximation does not take into account the depletion of the condensate caused by the interactions.
Additionally,   as the lattice depth $s$ decreases, the axial wave function of the initial condensate deviates from the Gaussian profile (\ref{initial}).

The wide spectrum peaks corresponding to the phase fluctuations  are shifted to the left relative to
the predictions of the model without the interactions.
In the case of $a=0$, the centers of the peaks in Figs. \ref{experiment_photos} and \ref{differences1} would  be  located at $kd/(2\pi)=0.5$ and $1.0$.
The model based on the Pitaevskii-Gross equation reproduces the peak shift well.
The broadening of the condensate, given by formula (\ref{broadening}), plays an important role in achieving the quantitative agreement.
Without this broadening,  when $\sigma=l_z$,  the shift is 2--3 times smaller.

\section{Conclusion}

\label{sec:Conclusion}

Within the framework of the Pitaevskii-Gross equation,  we  calculated the interference of a long chain of Bose condensates.
We identified two distinct interference regimes and their combination.
Comparison of the calculated interference fringe spectrum with the experimental data revealed  quantitative agreement  in the positions and widths of the peaks,  including the mean field shifts.
Regarding the heights of the peaks,  we found a discrepancy with some of the data.




\end{multicols}
\end{document}